# A Systematic Review of Human-AI Co-Creativity

SALONI SINGH, Vrije University Amsterdam, Netherlands
KOEN HINDRIKS, Vrije University Amsterdam, Netherlands
DIRK HEYLEN, University of Twente, Netherlands
KIM BARAKA, Vrije University Amsterdam, Netherlands

The co-creativity community is forging ahead with building more sophisticated and tailored systems to support and enhance human creativity. Design considerations or learnings from previous work can be an important and efficient starting point for this. To assist with that, we created a systematic literature review of 62 papers on co-creative systems. The selected papers span a diverse range of applications, from visual arts to design, writing, etc, where the AI's role goes beyond that of a tool- to engaging with the creative process or the user as an active collaborator. From the literature surveyed, we identified several dimensions of relevance to the design of such systems- (1) Phase of the creative process, (2) Creative task, (3) Proactive behaviour of the system, (4) User control, (5) System embodiment, (6) AI Model type. The review reveals that systems offering high user control give greater satisfaction, trust, and a sense of ownership over creative outputs, while proactive systems, when adaptive and context-sensitive, can deepen collaboration. 24 Design considerations were also gathered from the insights that emphasize the importance of encouraging users to externalize their thoughts and enhancing the AI's social presence and system transparency to build trust. Despite advancements, key limitations persist, including a lack of systems addressing early creative phases like problem clarification, and the complexity of user adaptation to AI tools.

---

[1]

---

[1]Contact Author : Saloni Singh, Vrije University Amsterdam

---

Authors' Contact Information: Saloni Singh, s.m.singh@vu.nl, Vrije University Amsterdam, Amsterdam, Netherlands; Koen Hindriks, k.v.hindriks@vu.nl, Vrije University Amsterdam, Amsterdam, Netherlands; Dirk Heylen, d.k.j.heylen@utwente.nl, University of Twente, Enschede, Netherlands; Kim Baraka, k.baraka@vu.nl, Vrije University Amsterdam, Amsterdam, Netherlands.

---

## 1 Introduction

The computational creativity community is increasingly exploring collaborative systems that are able to integrate human and AI expertise [99] Creative Support Tools (CSTs) [40] were already in use in the field of computational creativity even before the latest surge of AI and LLMs [47]. CSTs are designed to aid individuals or teams of people in the execution of creative tasks by either providing an environment to collaborate in, structuring the process, or providing resources, e.g. Pintrest [113], Grammarly [111], Dalle [110]. While CSTs significantly enhance the implementation phase of the creative process, they primarily act as tools. Since creativity is the generation of novel and valuable ideas or solutions [45] without necessitating the actual implementation of these ideas or solutions, and CSTs are built to support only the implementation phase of the creative process with the generation of the novelty and evaluation of the value of the ideas being done solely by the human users, therefore the AI system here is not deliberately contributing to the creativity. However, as these systems evolve, they are paving the way towards true co-creation where AI can strategically and in a contextually relevant manner contribute to the generation of novel and valuable ideas moving towards a more collaborative partnership.

Human-AI collaborative creativity [33] or just co-creativity thus involves two or more agents (of which at least one is human and at least one is an AI system) working together to generate novel and valuable ideas or solutions and may or may not implement them. In the context of human-AI collaborative creativity, "working together" requires bidirectional interaction with mutual adaptation. The collaboration must involve a shared goal with active contribution from the AI also such as providing meaningful suggestions, critiques, or ideas. Co-creative systems can take various forms such as AI-assisted design platforms or collaborative writing tools that suggest improvements or even new directions based on user input(e.g.- [35], [11], [85], etc.). To provide a clearer picture of what end-to-end co-creative systems can look like, we highlight three examples that illustrate the range of designs and interactions possible.

Lode Encoder [20], shown in Fig 1 first image, lets game designers create and edit levels through an innovative "painting" interface. It presents multiple suggestions for complete levels generated by the AI model after which the users can use "brushes" to "paint" sections of the different system-suggested designs onto the final level creator. Thus the system encourages designers to mix styles and explore new possibilities by offering tools like different brush sizes and an originality score to keep track of novelty in design creation. This system further challenges the creativity of the users by not offering traditional editing tools and forcing them to work within the confines of the AI-generated levels. It also offers them grounds to test the level to ensure playability. Another example is a co-creative musical system [104] (NOISA), second image in Fig 1, that monitors performers' engagement during improvisational performances and intervenes with subtle, contextually coherent melodies when engagement levels drop, helping sustain the user's creative flow throughout the performance. Lastly, Cobbie [66], shown in Fig 1 bottom image, an embodied robotic system, collaborates with users in sketching on paper, by generating ideas that are meant to cause a conceptual shift in the thinking of the user. While Cobbie provides intelligent suggestions, the user retains full control by controlling when to give the pen to Cobbie and selecting the area it will draw in. The user is also engaged by the robot's engaging movements and sound feedback to emulate social interaction. These systems illustrate the broad range of applications and interaction dynamics present in co-creative tools.

These systems facilitate a more dynamic and interactive creative process, letting individuals explore ideas they may not have considered independently. Building AI systems that act more as collaborators or colleagues than tools can help us fully utilize the potential of AI [73]. By progressing to assistants or co-workers, AI can handle routine or complex tasks, allowing humans to focus on higher-order creative thinking. Collaborative creativity can create ideas or results that are more creative than each individual could have achieved alone [98].

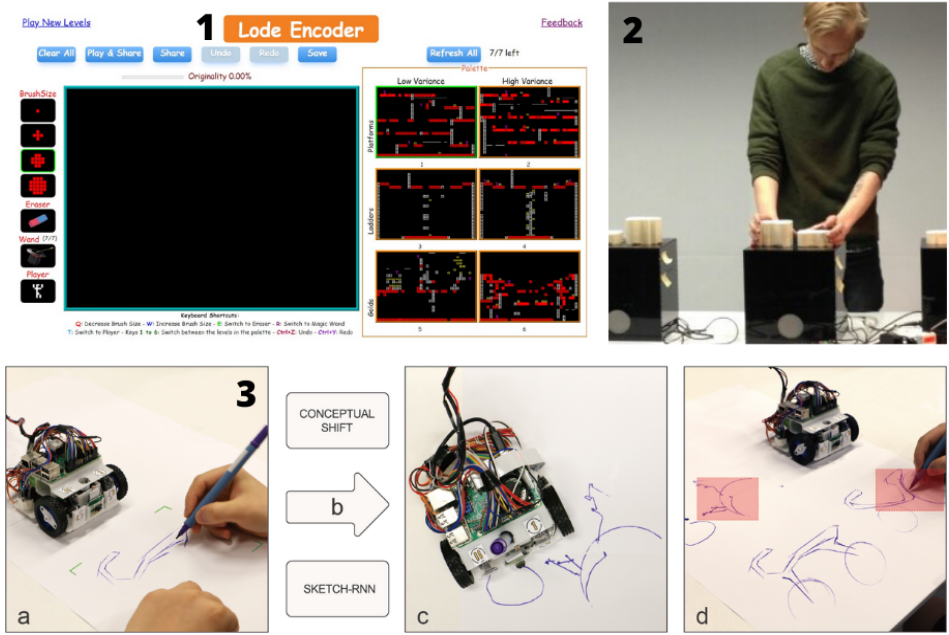

Fig. 1. Examples of Co-Creative Systems. 1.LodeEncoder [20] 2.NOISA [104] 3.Cobbie [66]

As the co-creativity community forges ahead with building more sophisticated and tailored systems to support and enhance human creativity, design considerations or learnings from previous work can serve as a valuable and effective foundation. There are a number of works in this space, but our approach offers a unique perspective as shown below. Alsharhan [9] and Bossema et al. [23] have done valuable work in synthesizing the trends in structure and function design in their respective creative domain/task of natural language generation and art-based therapy respectively. To build upon existing research but also ensure the broader applicability of these considerations to more domains, this review chose a scope comprising of all artistic tasks, thus working with a broader scope than [9] and [23]. Rezwana and Maher [94] aimed to propose a framework for categorizing existing hybrid creative systems based on their interaction dynamics. Our aim, on the other hand, delves into trends in literature, user preferences, insights, and lessons from existing user studies. This allowed us to offer a qualitative analysis of how these dimensions impact users and the creative process. Hwang [48] investigate how the creative systems, in their scope, support the different stages in the human creative processes. Similarly, we tracked the phases of the creative process for the same purpose but focused exclusively on co-creative systems to identify trends specific to this field, whereas Hwang's work examines CSTs. The interdisciplinary work by Mccormack and his team [80], studied the various perspectives of successful human collaborations and hypothesized possible design considerations for collaborations with an AI system. This review takes their work forward by compiling real-world insights, benefits, and findings from existing systems and user studies that address user feedback within the context of the mentioned considerations.

This review builds on these works to link design decisions, in co-creative studies, to impacts on creativity and user engagement, to provide evidence-based recommendations for creating systems that effectively stimulate and support human creativity. This review also covers the years 2015-2023, thus incorporating the recent introduction of LLMs into computational creativity and its effects on the field of Co-Creativity. Now that AI models generate art, compose music, and even write poetry, pushing the boundaries of what machines can achieve [107], we look into how one may best harness these capabilities of the AI systems, within the creative environments in which they operate.

In this article we present an analysis of the body of work (62 papers) we have systematically gathered based on PRISMA guidelines. The papers were selected from three major databases: IEEE Xplore, ACM Digital Library, and Scopus, using a combination of keyword searches related to AI, collaboration and creativity. The inclusion criteria required that papers focus on systems where AI actively and intentionally contributed to creative processes alongside human participants. We restricted the scope to artistic tasks—such as visual arts, music, design, and performance—where the end goals were dynamic and evolved throughout the creative process. Exclusion criteria eliminated tasks with predefined end goals or deterministic outputs, such as academic writing, software development, or technical problem-solving, where creativity was not a core focus. Additionally, only papers that evaluated the system through user studies or qualitative assessments were included. After a rigorous screening and review process, we distilled the findings into system and study design considerations. These considerations are organized into dimensions that arose organically from the literature and inspired by the frameworks [94],[48] laid out in previous reviews. The final set of dimensions (Fig. 2) that capture a non-exhaustive set of variations in the design of Human-AI co-creativity systems are:

- Creative Task (Section 4.1): Creative task evaluated in the user studies.
- Phase of the Creative Process (Section 4.2): The 4 stages of the creative thinking process according to Osborn and Wallas
- Proactive System Behaviour (Section 4.3): An AI system's ability to self-initiate the process of identifying potential challenges or opportunities, predicting future outcomes based on its inputs, and taking actions to influence its environment or adjust its operations to achieve specific goals or avoid problems.
- User Control (Section 4.4): The level of control over the AI system.
- System Embodiment (Section 4.5): Categorization of the embodiment of a system, ranging from no embodiment to physically or virtually embodied.
- Model Type Categorization (Section 4.6): Category of AI models used in the systems.

Some user evaluations and insights, from the subjective assessments (e.g., perceived creativity, engagement) emerged from the literature beyond the defined dimensions. We use the conceptual frameworks and evaluation strategies proposed by Mccormack et al. [80] and [27] as guides for categorizing them into the following evaluation themes:

- User Perceptions of System intent (Section 5.1)
- Roles and Expectations (Section 5.2)
- System Transparency and User Guidance (Section 5.3)
- System Consistency (Section 5.4)

The structure of this review is organized as follows. Section 2 (Methodology) details the systematic approach used to identify, screen, and select the papers included in this review, including the criteria for inclusion and exclusion, as well as the systematic categorization of papers using a predefined set of dimensions to facilitate structured analysis. In Section 3 (Dimensions Overview), we define key concepts and terminologies central to the review, such as proactivity and the stages of the creative process among others. In Section 4 (Analysis), we present the dimensions that emerged from the literature providing a detailed analysis of trends, correlations and design considerations that emerged. Section 5 (Evaluation Themes) organizes the findings from the user studies, highlighting preferences, design considerations(named DC followed by a number, collected into a table in Appendix A 3), and challenges, including insights into user trust, role expectations, and consistency in AI behaviour. Finally, in Section 6 (Conclusion), we summarize the primary findings of the review, discuss their implications for future research and design in human-AI co-creativity systems, and suggest directions for addressing gaps in the current body of work.

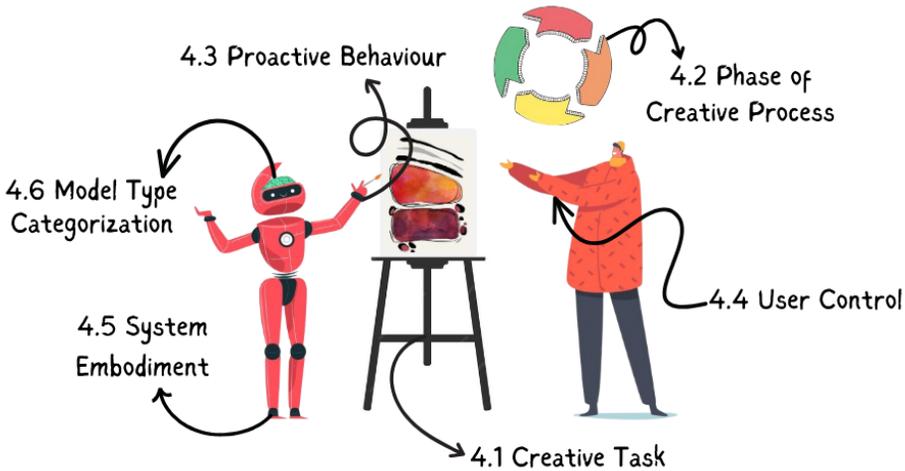

Fig. 2. Six Dimensions Illustrating Differences in Key Design Aspects for Co-Creative Systems

## 2 Methodology

In this section, we first touch upon what is creativity and co-creativity and how they guided the inclusion/exclusion criteria for the review. Then we expand on the methodology of the collection of the papers.

### 2.1 Initial Selection Criteria

Our selection criteria were guided by literature on (co-)creativity. Newell, Simon, and Shaw [83] propose that creativity or creative thinking is essentially a specialized form of problem-solving behaviour that emerges under specific conditions. The outcome of the process must be new and valuable, the process should involve unconventional thinking and high levels of motivation and persistence, and finally, the problem is typically vague or poorly defined at the outset. A critical part of the creative process, therefore, involves not only finding a solution or outcome but also clearly formulating what the problem actually is. Thus co-creativity [33] would involve two or more agents working together to generate novel and valuable ideas or solutions. In the context of this review on Human AI co-creativity, at least one of the agents is human and at least one is an AI system.

These definitions guided our inclusion-exclusion criteria as follows. We only included systems where both the AI and the Human intentionally contributed to the creative part of the task. We also restricted the definition of creative tasks to artistic tasks with dynamically shifting end goals according to the criteria of Newell et al. [83] in their definition of creative. Therefore, tasks like academic paper writing, code development, etc. were excluded from the scope. Finally, we restricted the inclusion criteria of "AI" to any computational system that exhibits behaviour that is not solely dictated by hard-coded rules. This includes systems that utilize machine learning, probabilistic reasoning, or other adaptive techniques to process inputs, learn from data, and make decisions or generate outputs in ways that are not explicitly pre-programmed by developers. The decision to adopt this definition stems from the desire to focus on systems that exhibit a degree of autonomy and adaptability, qualities that distinguish AI from traditional software that follows deterministic, rule-based logic.

## 2.2 Selection Procedure

The methodology for selecting papers for this systematic review was designed to ensure a comprehensive and relevant dataset. Three primary databases were chosen for the literature search: IEEE Xplore [112], ACM Digital Library [1], and Scopus [114]; for their extensive coverage of high-quality, peer-reviewed content in engineering, computer science, and interdisciplinary research. The search across these databases was last conducted on January 30th, 2024. The search strategy was a combination of keywords and Boolean operators, targeting both the document titles and abstracts of potential studies. Here the * meant wildcard/s match, and anything inside " " was an exact string match. The space before the search word art was added to not match with the words "state-of-the-art". This helped to identify papers that explicitly discussed the collaborative interactions between humans and AI within creative contexts.

| Creative | AI | Collaborate |
|---|---|---|
| " Art" | "Agent" | "Co" |
| " Artist" | "Robot" | Collaborat* |
| Creativ* | "AI" | "Mixed Initiat*" |
| Sketch* | "Machine Learning" | Interact* |
| "Writing" | "Computational" | |
| "Writer" | | |

Table 1. Keywords Searched in Databases

The initial search resulted in a total of 2,668 papers across the three databases after filtering on the year of publication 2015-2023 and language as English: ACM Digital Library: 1,036 papers, Scopus: 1,228 papers, IEEE Xplore: 404 papers. The data collection process adhered strictly to the PRISMA (Preferred Reporting Items for Systematic Reviews and Meta-Analyses) guidelines [86] to ensure a rigorous and systematic approach. Following the initial import into Rayyan [6], the dataset underwent a thorough cleaning process. Duplicates were identified and removed (using Rayyan's automated feature [3]), resulting in the exclusion of 538 duplicate entries (2130 papers left). After filtering out short papers (fewer than five pages) (379 papers) and non-journal/conference papers (19 papers), the dataset was reduced to 1732 papers. The first phase of screening involved an initial review of titles and abstracts by a single reviewer. This phase focused on excluding papers that did not demonstrate a clear open-ended creative task. After this initial screening, 1110 papers were excluded, leaving 622 papers for further review.

A second, more rigorous screening was conducted by the four authors, with all the papers being reviewed by the lead reviewer, and at least one other reviewer independently blind assessing to each paper. Any discrepancies between reviewers were discussed and resolved by a third reviewer, and where necessary, automation tools (e.g., Rayyan's conflict resolution feature [5], etc.) were utilized to streamline the process. After this round of screening, 368 papers were excluded, resulting in 254 papers for the full-text review section. In addition to the structured search through the selected databases, papers from the Computational Creativity Conference (computationalcreativity.net [2]) were manually curated due to their high relevance and absence from traditional databases. All accepted papers from the conference from 2015 to 2023 were reviewed for the same inclusion keywords, filter criteria, and abstract screening and finally integrated into the pool of studies for the full paper screening round (41 papers).

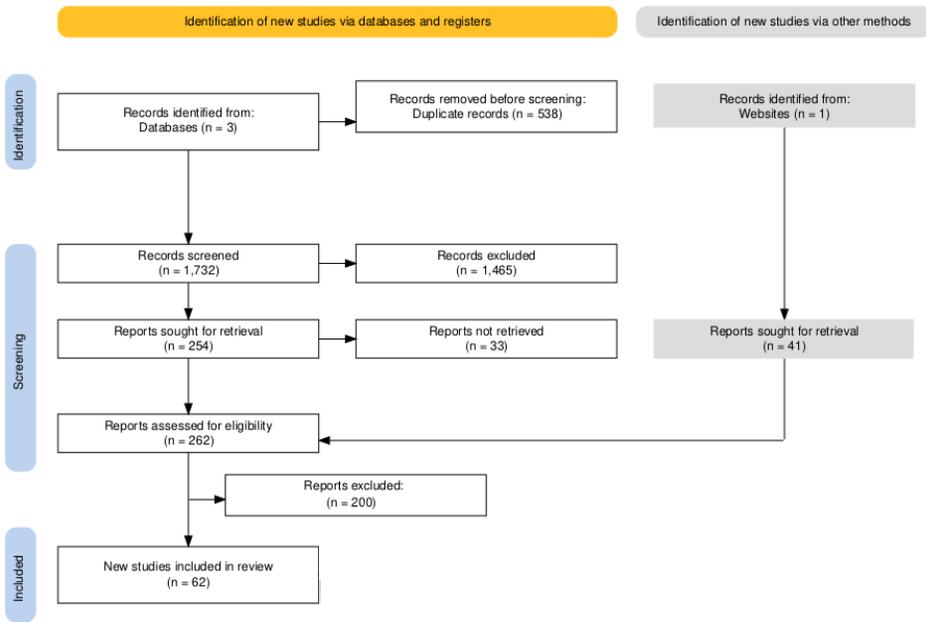

Fig. 3. Screening Process In Line with PRISMA guidlines

The final phase involved a full-text review of the 295 papers (41 computationalcreativity.net [2], 254 from the databases), conducted by the lead reviewer, with any conflicts or uncertainties resolved through discussion with the team. The inclusion criteria at this stage emphasized the significance of the AI component contributing to the creative process and the scope of AI models to be included. Papers without any results or evaluation were excluded. There was another round of manual duplicate removal at this stage. After the completion of all screening and review processes, the final set of papers included in the review totalled to **62** papers. These papers collectively contribute to a nuanced understanding of the current state and future directions of human-AI collaboration in creative endeavours.

## 2.3 Paper Categorization

In order to systematically examine the systems included in our review, we assigned specific categories to capture the dimensions shown in Fig2. The selection of these labels was a deliberate and iterative process. The final set of dimensions were: the proactive behaviour of the AI, User Control, Creative Task, System Embodiment, AI Model Type, Publication Year, and Phase of the Creative Process. Each of these dimension categories had some possible sub-categorizations that were defined and refined through a collaborative review process involving all team members. Once all the possible sub-categories per dimensional category were finalized, the first author manually scanned each paper and applied them. This process allowed us to structure our analysis and categorize the data systematically. Unlike dimensions, the papers were not tagged by the categories of evaluation themes they talked about. The evaluation themes are just general considerations and learnings to keep in mind when designing a co-creative system, outside the interaction and system design space. The subsequent data cleaning and plotting provided a clear visualization of the trends and patterns that emerged from our review. The definitions and scope of each dimensional category along with each possible sub-categorization under that dimension and the rationale for their selection, are provided in the corresponding subsections under Section 3. A count of the number of papers published per year in the final set of papers in the scope of this review is shown in Fig 4. The upward trend in the number of papers in this field over the course of 2015-2023 indicates growing interest in the co-creativity field.

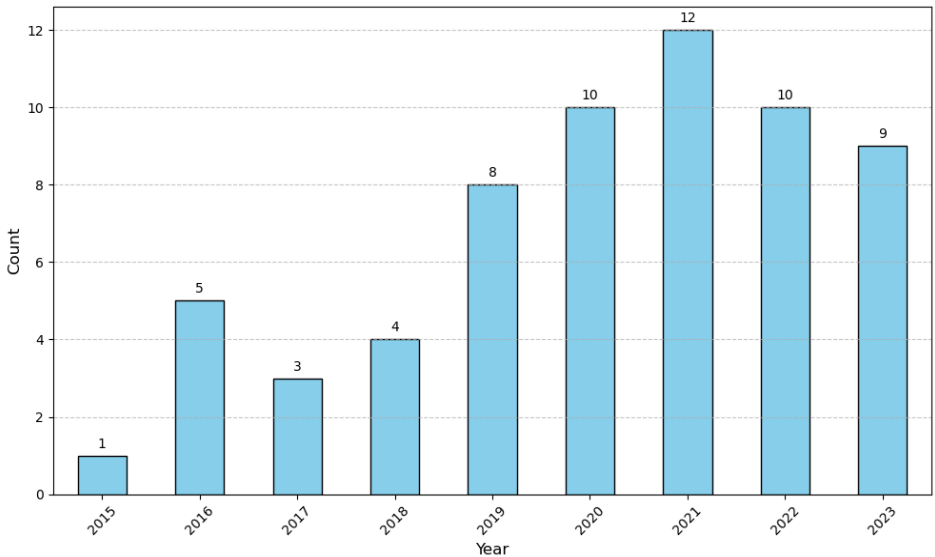

Fig. 4. The annual count of papers included in the final set of this review, covering the years 2015–2023. The upward trend in publications over this period highlights the increasing research interest and engagement in the field of co-creativity.

## 3 Dimensions overview

In this section, we define and provide relevant background on each dimension of Fig. 2. This is to provide a structured foundation for understanding how the identified themes and dimensions interrelate.

### 3.1 Creative Task

This dimension captures the creative outcome of the tasks evaluated in the user studies. It was meant to capture the artistic domain to track how preferences or trends change based on task type. Zhang's paper [122] laid out a taxonomy of creative tasks people perform in their everyday life and work, which we adjusted to artistic tasks only, to use as the "Creative Task" dimension. Originally Zhang's taxonomy of creative tasks categorizes various activities across different domains of everyday life and work. The tasks covered multiple domains, including Visual Arts (e.g., painting, interior design), Performing Arts (e.g., filmmaking), Music, Literature (fiction and nonfiction writing), Arts and Crafts (e.g., furniture making, jewelry design), Cooking, Science and Engineering (e.g., problem-solving, programming), and Everyday Tasks (e.g., learning new skills, solving non-technical problems). This taxonomy captures a wide variety of creative activities, offering a detailed framework for analyzing trends and preferences based on the type of creative task performed. A slightly adjusted version of this taxonomy was used for the papers in our review. Science and Engineering was renamed to Design and engineering, and architecture design was moved from visual arts to design and engineering since it was not as open-ended as the other visual arts. performing arts included improvisational games, theatre, dance, and live storytelling. Finally, game design was not listed in the taxonomy but cropped up a lot in the papers under the scope of the study, so it got its own category.

### 3.2 Phase of the Creative Process

The phases of the creative process dimension capture the stages through which the AI system meaningfully contributes to or coaches humans. Understanding this is essential for evaluating the system's effectiveness and identifying areas for improvement. It allowed us to identify which phases of creativity are less targeted by existing systems, potentially highlighting niches that future projects could address. The study of creative processes has led to the development of various theories [72], where among the most notable theories is the Creative Problem-Solving (CPS) Model by Osborn-Parnes [87]. It is particularly well-suited for integrating AI as a co-creator, especially in the context of open-ended artistic processes, due to its structured yet flexible approach, focusing on the set of actions taking place, unlike models such as Graham Wallas' Four-Stage Model [105] or Lubart's 7C's model [74], which emphasizes subconscious processes like incubation (a period of stepping away from a problem, allowing ideas to develop subconsciously) or illumination (the "aha!" moment when a new idea or solution suddenly emerges) that are inherently human and hence not directly applicable to human-AI settings.

The 4-step CPS model [87] breaks down creativity into four stages— *Clarification*, *Ideation*, *Development*, and *Implementation*. *Clarification* is where the problem or challenge is thoroughly understood and clearly defined through extensive research and information gathering. For example, an artist might explore themes and cultural contexts to refine the conceptual basis of their work. Once the problem is defined, the process moves to *Ideation*, where a broad spectrum of potential solutions is generated through techniques like brainstorming and mind mapping, encouraging creativity without immediate judgment. In the *Development* stage, these ideas are critically evaluated, refined, and transformed into practical solutions. Finally, in the *Implementation* stage, the refined solution is executed. This dimension can have multiple stages tagged per paper since they might target multiple phases of the creative process. For example, "improvisation" tasks were typically tagged under both the "ideation" and "implementation" phases due to their nature of generating and executing ideas in real-time. While these stages are presented as a linear process, in reality, they often form a cyclic or iterative pattern, with dynamic movement between stages as needed. For the sake of simplicity in categorizing these stages for study within the human-AI co-creative setting, we will treat them as distinct phases, acknowledging the fluidity and overlap that naturally occurs in creative endeavours.

### 3.3 Proactive System Behaviour

When we discuss proactivity here, we refer to the system's proactivity. Our definition was inspired by Parker and Bindl [21] who defined it as "self-initiated and future-oriented action that aims to change and improve the situation or oneself". They expand on the definition with: "First, proactive behaviour is self-initiated, which means that this behaviour is enacted without being told to or without requiring an explicit instruction". In the context of human-AI systems, initiative-taking can be observed when either the human or the AI agent takes control of the process by taking unprompted action, proposing steps, generating ideas, or directing the focus of the interaction. While it is possible to have multiple dialogues/actions within various subgroups, as mentioned in the paper [30], we assume only one dialogue/action at any given time for simplicity. This assumption holds because systems in this review's scope typically involve just one or two agents interacting simultaneously. Second, proactive behaviour is future-focused, meaning it addresses anticipated problems or opportunities that extend beyond the immediate next action or current requirements within the creative task or activity, focusing instead on future needs or goals that emerge over the course of the work. Third, proactive behaviour is "change-oriented, involving not just reacting to a situation but being prepared to change that situation or oneself in order to bring about a different future." This definition which originally came from the social sciences/psychology fields has now been co-opted into human-robot interaction literature too [25], [90].

Proactivity was an interesting dimension in our analysis because it can help get a system closer to the AI-as-teammate dynamic. By demonstrating the ability to suggest or set goals, an AI moves from being a passive assistant to an active collaborator. Proactivity mirrors an important dynamic in human-human collaboration, where negotiation of goals and mutual contributions often define successful teamwork [97]. We identified five possible sub-categories in this category, adapted from Buyukgoz [25]:

- **Anticipating User Needs** The system perceives the user's needs and offers support proactively, where the system may act to provide suggestions without explicit prompting.
- **Information-Seeking** The system takes proactive steps to improve its knowledge by seeking information from the user or other sources. This could involve the system asking for validation, clarifying uncertainties, or identifying gaps in its knowledge base. This behaviour is useful for systems that need to refine their understanding of the task or environment continuously, such as artistic tasks with emergent or shifting goals.
- **Anticipating Goal Needs** Originally termed "anticipating possible plan failure and plan repair" by Buyukgoz [25], it involves the system foreseeing potential obstacles or failures in the current plan or process and proactively taking steps to prevent or repair these issues.
- **Opportunity-Seeking** This subcategory was introduced by us to note behaviours of the system where it may identify and pursue new goals or opportunities that the user did not explicitly set. Unlike "anticipating goal needs," where the system optimizes a predefined goal(either by the nature of the task or user-set goals), opportunity-seeking behaviour involves the system suggesting or initiating new goals or paths that could potentially add to the creative process. In the context of creative tasks, this proactive adjustment or addition to the existing goals often represents the collaborative nature of co-creative systems, making this an important feature to track. This category was absent from the Buyukgoz [25] paper since their focus was on robots helping humans achieve their vision by "trying to converge toward the needs of the goal". Therefore the robot in this situation does not do its own goal setting or suggesting, causing this category of proactive behaviour to be absent from the list of proactivity types in Buyukgoz's paper while we consider a broader scope of proactivity.
- **Not Proactive** This subcategory is for papers where the system did not show any kind of proactive behaviours.

Tracking these types of proactivity is valuable because each dimension captures a distinct way the AI can contribute to the user's workflow and creative process. Given that a system can exhibit multiple types of proactivity, multiple sub-categorizations, within the proactivity dimension, are possible per paper (except for "Not Proactive").

## 3.4 User Control

This dimension captures the level of control humans have over the actions and contributions of the AI system in the final product resulting from the collaborative process. This dimension is adapted from User Experience (UX) Design. In UX design, user control is defined as the degree to which users can direct, influence, or override the system's actions to achieve their goals [89]. UX research has shown that "users who feel in control are more likely to actively use and enjoy the experience" and gain a greater sense of agency [46]. This dimension is important to track the user preferences for control over the AI contributions. It would help understand users' perceptions of seeing the AI as an equal collaborator vs a tool for generating ideas, suggestions, or critiques. It reveals how much involvement users want from the AI and how much control they prefer to retain in the creative process. Three subcategories constituting the three levels of user control were identified:

- **"No control"** Once the AI contributes to the piece, whether it's a canvas with a painting, a cooking recipe, or an improvisational performance, it cannot be altered or removed by the human.
- **"Limited control"** While the human has some degree of influence over the direction of the final product, their choices are limited by boundaries set by the AI. In this case, the AI maintains some control over the process and its direction. For example, in a co-writing task (e.g.[84]), the AI might offer five possible directions for the story based on what has been written so far. The human can only choose from these options, meaning the AI controls the scope of potential outcomes, thereby influencing the narrative's direction.
- **"Full control"** The human can remove or alter the AI's inputs, effectively giving them full control over the outcome.

## 3.5 System Embodiment

System embodiment dimension refers to the physical representation of AI, whether through avatars, robots, or interfaces. Tracking embodiment across systems can provide insights into how physical representation impacts task engagement, especially in creative fields like design and performing arts. For this review, the AI system in a paper could be categorized under one of the following options - Embodied (physically embodied like a robot or some other dynamic interface like Cobbie [66]), Virtually Embodied (the AI system is given a virtual/non-tangible body like LuminAI [115] ), Non-embodied (No body given to the AI system like ChatGPT [109] or Grammarly [111]).

## 3.6 AI Model Type

We defined this dimension for this paper, to track how the introduction of LLMs affected the co-creativity field. Therefore the dimension has these subcategories available -

- LLM: Large Language models/ transformer architecture-based models
- NN: Neural Networks that don't fall under the transformer architecture-based models mentioned above
- Other: No neural networks were used in the creativity part of the model. E.g.- k-nearest neighbors (KNN) algorithm, cooperative contextual bandit algorithm, etc.

## 4 Analysis

In the following sections, we analyse the distribution of the surveyed works across these six dimensions and examine any relevant interrelations between different dimensions. Where applicable, we additionally select noteworthy findings from the evaluations and user studies of the papers for each dimension and provide preliminary design considerations in connection to these findings. Table 2 provides a summary of categories and subcategories for easy referencing.

| Creative Task | Representative Papers |
|---|---|
| Music | [79], [24], [28] |
| Visual Arts | [26], [57], [92], [103] |
| Design/Engg. | [14], [19], [64] |
| Literature | [39], [119], [41] |
| Game Design | [17], [11], [100] |
| Performance arts | [52], [62], [61] |
| Misc | [49], [37] |
| **Phase of Creative Process** | |
| Clarification | [49], [16] |
| Ideation | [38], [70], [8], [120] |
| Development | [54], [50], [56] |
| Implementation | [116], [115], [104] |
| **Proactive Behaviour of AI** | |
| Anticipating user needs | [24], [20], [67] |
| Anticipating Goal needs | [64], [77], [10] |
| Opportunity seeking behaviour | [75], [104], [85] |
| Information seeking behaviour | [57], [16], [104] |
| No proactivity | [66], [18], [43] |
| **User Control** | |
| Full control | [16], [54], [56] |
| Limited control | [54], [38], [70], [81] |
| No control | [8], [108], [104], [118] |
| **System Embodiment** | |
| Embodied | [38], [8], [26] |
| Virtually Embodied | [115], [52] |
| Non Embodied | [70], [56], [116], [93] |
| **Model Type Categorization** | |
| NN | [16], [54], [119], [44] |
| LLM | [70], [38], [14] |
| Other | [116], [102], [49], [115] |

Table 2. Dimensions for Key Design Aspects for Co-Creative Systems with Subcategories and their Representative Papers

## 4.1 Creative Task

Figure 5 shows visual arts and literature to be in the lead, with a rise in such systems from 2019 onwards. Although this aligns with the timeline of the GPT and LLM releases, the usage of LLMs in the papers alone does not account for the rise in the number of papers published. They were already on the rise. Another noteworthy finding is that most LLMs are used in literature, while neural networks are predominantly applied in visual arts 6.

Music tasks often employ models categorized as "Other", such as Random Forests in [28] to generate melodies. However, some use neural networks (NNs), like temporal CNNs employed in [79] to create dynamic musical compositions. Visual arts use diverse AI models. Non-neural network approaches include cooperative contextual bandits, as in [58] and [57], for image-based contextual exploration. Other tasks utilize NNs like SketchRNN, CNNs, and GANs. For example, CNNs are used in [55] to find visually and semantically similar images to provide inspiration to the user, while GANs feature in [8] in a turn-based sketch completion game. Hybrid techniques such as genetic algorithms [19] for generating design sketch alternatives and LSTM-k-means-word embedding combinations in [56], for another version the system that finds visually and semantically similar images to provide inspiration to the user, are also notable. Combining LLMs with GANs, [26] explores speech-to-image synthesis for interactive artistic creation. Finally, we observed that many sketch-based tasks (even outside visual arts tasks) utilized the Quick, Draw![4] dataset, emphasizing its value as a foundational resource for creative sketching systems. Performing arts predominantly use NNs or non-NN techniques, with few LLM applications. For example, VAEs combined with rule-based systems guide improvisational decision-making in [52]. Non-NN models like YOLO[13] are used in [12] to generate improvisational actions. An exception, [78], employs tensor2tensor (an LLM) alongside deep neural networks for generating dialogues for theatre contexts.

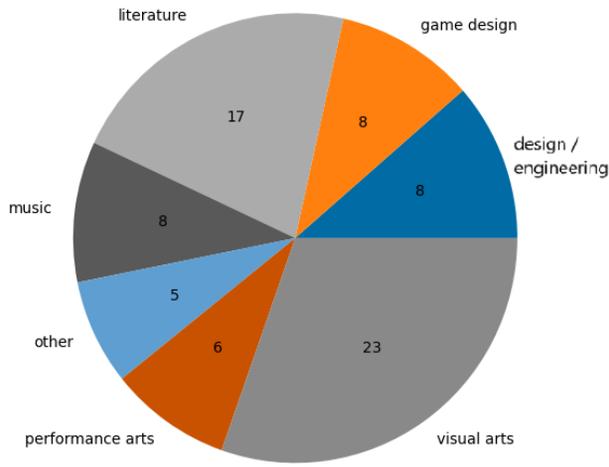

Fig. 5. Creative Task-Wise Classification of Co-Creative Systems in Surveyed Papers (2015–2023)

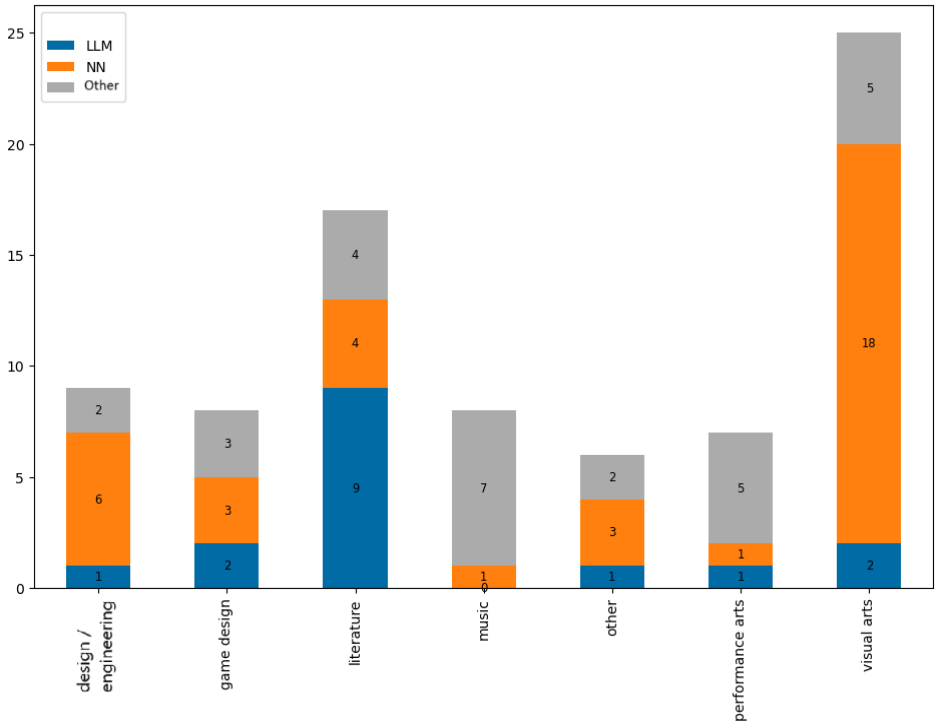

Fig. 6. The relationship between different creative tasks and the types of models used.

Game design evenly distributes its reliance across NN, LLM, and non-NN models. For instance, [43] uses Markov chains, LSTMs, and Bayesian networks as three alternative ways of procedural game generation in the same system. Genetic algorithms in [10] generate game "rooms" tailored to specific fitness functions. LLM-based systems like GPT-2 drive quest generation in [17], while AutoGPT generates board games in [100] with an online survey of exiting board games and feedback from the users playing as game testers. Design and engineering tasks primarily utilize NNs, with occasional use of LLMs and non-NN models. [14] use LLMs by integrating natural language prompts with Sketch2Code for rapid iteration in creative coding. Neural network models such as RNNs and CNN-LSTMs support sketch-based ideation, as seen in [66] and [53]. Genetic algorithms also appear in [64] for configuration design. Literature tasks are split across LLMs (50%), NNs (25%), and non-NN models (25%). Random Forests guide lyric creation in [28]. LLMs like GPT-2/3 dominate in tasks such as narrative refinement ([70]) or storytelling ([38]). Non-NN NLP techniques, like n-grams in [22], encourage linguistic creativity, while RNNs assist in story writing, as seen in [108]. By examining these task-specific applications, we see how AI not only complements human creativity but also shapes the creative process itself, highlighting the importance of tailoring AI techniques to the unique demands of each domain.

However, a possible design consideration(DC) can also be to build AI systems that break beyond the boundaries of a single creative domain. Rather than restricting AI outputs to the established conventions of one field, cross-domain inspirations can be intentionally integrated to enhance the creative process. For example, in literature, AI could incorporate visual metaphors or artistic elements from visual arts datasets, offering a fresh, painter's perspective on narrative structure and emotional tone. In visual arts, AI could draw on narrative techniques—such as developing characters or settings—to guide the creation of more dynamic or abstract compositions. This approach can **DC 1:** *encourage users to explore unconventional connections between creative domains and develop hybrid forms of expression, pushing beyond the patterns that exist within any single domain* .

## 4.2 Phase of Creative Process

The review classified systems based on the phases of the creative process they supported, with 5 papers focusing on clarification, 49 papers addressing ideation, 24 papers emphasizing development, and 32 papers supporting implementation. The clarification phase has the lowest representation, indicating that few systems focus on this early stage of the creative process. Development and implementation phases are moderately represented, but ideation stands out with the highest number of papers (49 out of 62 papers) by a significant margin. When examining Figure 12, we see that visual arts and literature have a focus on the ideation phase. Music and performing arts show low representation in the development phase, because most systems in these domains are designed for improvisational tasks rather than structured development. Design and Engineering systems predominantly feature ideation and development.

The development phase has been increasingly targeted since 2018 and appears to be on a general upward trend per Fig 11. Ideation, on the other hand, saw a significant increase starting in 2019. This again may look related to the LLMs development timeline, and while it does hold true for 2022 onwards, the increase from 2019 was unrelated to LLMs. So there does seem to be an increasing interest in targeting the development phase of a creative task with co-creative systems. Finally, the clarification and development phase had almost no embodied systems, this is also due to the fact that most embodied systems were improvisation-based tasks, which were therefore under the ideation and implementation phase of creativity.

We look at some examples that illustrate how various AI techniques and models are applied across different phases of the creative process to better understand the diverse roles AI plays here. During the clarification phase, [16] applies libraries such as SpaCy to generate insightful questions based on user input to prompt reflection and look for clarity in writing. Creative wand uses LLMs for the same reflective questions too [67]. [31] uses Textrank [82], Golve embeddings [91] and RoBerta [68] to generate text summaries of the user's input to serve as the conceptual exploration in the reflection and clarification phase of the creative process. [58] also uses the cooperative contextual bandit Machine learning model to prompt reflection in the human by showing abstract notions related to the perceived mood of the existing board of images collected by the user.

In the ideation phase, different AI techniques/models are employed to support the divergent generation of ideas with LLMs accounting for around 20% and NN, "Other" models around 40% each. For instance, StoryDrawer[119] leverages a neural network-based system, SketchRNN, for both classifying and generating sketches, helping children in storytelling by generating visuals for the story. For generating dialogues and quests in role-playing games, a combination of GPT-2 (restricted to knowledge graph nodes) and knowledge graphs is used in Personalized Quest and Dialogue Generation[17]. It guides the generation process with specific grammar and rules, showing a large language model's flexibility. [39] used a task-specific dataset and rule-based behaviour designed to encourage children with idea generation as well as offering ideas from the system's side.

In the development phase, the Model types were relatively evenly distributed at 30-40% each. systems like Picture This[92] and [69] use generative adversarial networks (GANs), a type of NN, to merge or generate images, aiding users in refining or combining ideas to make them stronger. [100] uses AutoGPT to create games based on existing game designs online within constraints set by the users, thus going through the ideation process itself. [41] has GPT3 offering alternate endings for the user to choose between for the continued development of the story. Finally, [49] uses a task-specific dataset and rule-based behaviour to offer different ingredients and techniques to users to help them refine a dish recipe.

Finally, in the implementation phase with LLMs being used the least at 13%, Dream painter[26] explores creative possibilities through speech-to-image synthesis, employing systems like CLIP Draw(LLM) combined with rule-based algorithms to generate artistic drawings. [104] used ML models to gauge the engagement of the users to choose when to step in and make changes to the musical output to keep them engaged. [108] uses various NLP techniques and RNN models to add its ideas and often finishing touches to some changes suggested by the users, to the final version of the story-telling game. These examples demonstrate how diverse AI techniques are employed across different phases of the creative process, with each phase benefiting from tailored approaches that leverage the strengths of specific models and methods.

*Noteworthy findings.* Karimi et al. [54] concluded from their user study that a low novelty mode helped designers add more details to the drawing, thus helping with the implementation part of the "finishing touches", while high novelty mode was more useful for transforming or adding new features- the development/ideation phase. This shows that small design/ algorithm choices can make a marked difference in the way the system is used. However, it also provides an interesting insight into how to affect convergent and divergent thinking phases of the creative process. Specifically, systems designed to target a particular phase should account for this: a high novelty factor is more suitable for fostering divergent thinking, while a low novelty factor is better suited for supporting convergent thinking.

Rafner and his team [92] saw that the image generation tool, though not initially intended for the clarification phase, forced people to visualize and reflect on their vision better because the prompt needed more details. In [32] they had a similar discovery that switching from diegetic writing to non-diegetic instructing shifted the writers' focus from the narrative of the story to the prompt engineering, inadvertently making them reflect on their ideas/story. [31] system was designed with the intended use of AI summaries of the participant's works for reflection. While it did work for the intended purpose they noted that the AI summaries were instead used as scaffolding for self-reflection. Users compared a "mental summary" to the generated summary. So the AI annotations didn't replace the mental summarization process, but instead just empowered it.

**Design considerations**. Some design considerations that emerge from these papers and their extensive evaluations for systems to support creativity and reflection are given below. It's important to strike the right balance between automation and active user engagement. Automation should make tasks easier, but users still need to stay mentally involved—this is where real reflection and growth happen.  *DC 2: Introduce task shifts to help users see their work from fresh angles and spark deeper thinking* [71], for example by moving between storytelling and crafting prompts. *DC 3: Encourage users to externalize their thoughts to make reflection a central part of the creative process* [42]—whether through detailed inputs or comparing their ideas with AI outputs.  *DC 4: Create moments of productive tension to push users to engage more deeply and refine their creative ideas* [88], like reconciling a mental summary of their writing with an AI-generated one.

### 4.3 Proactive System Behaviour

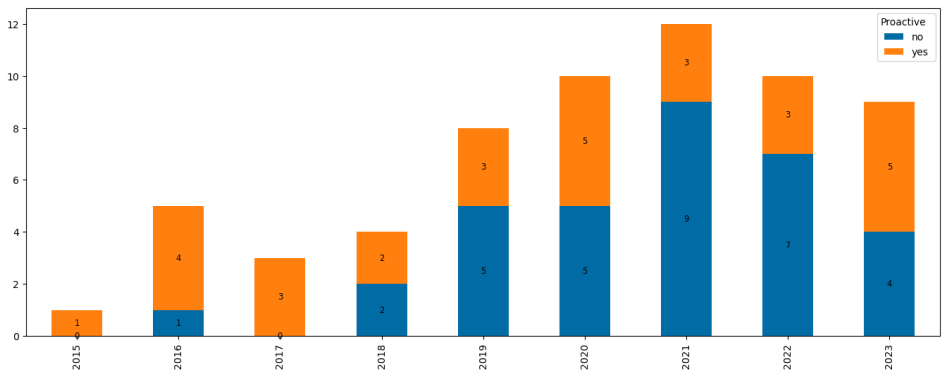

Fig. 7. The distribution of proactive and non-proactive systems across publication years. Systems exhibiting any form of proactive behaviour—including opportunity-seeking, information-seeking, anticipating goal needs, and anticipating user needs—are grouped under "Yes," while non-proactive systems are categorized as "No."

The review subcategorized systems based on types of proactivity, with 20 papers focusing on anticipating user needs, 3 papers addressing information-seeking behaviour, 17 papers highlighting anticipating goal needs, 23 papers examining opportunity-seeking behaviour, and 33 papers that were not proactive. Of the 62 papers, there were 29 papers that show at least one subcategory of proactive behaviour and 33 non-proactive papers. The distribution of proactive systems over the years in Fig 7 shows that the number of systems labelled as "Yes"(for this figure opportunity seeking, information seeking, anticipating goal needs and anticipating user needs were combined into "yes" and duplicate entries of the same paper were removed to get an accurate count of how many shows any signs of proactive behaviour) for proactivity has remained more or less constant, despite the significant increase in co-creative systems overall. At the same time, the number of systems labelled as "No" for proactivity has increased. However, it seems reasonable to say for now, that interest in proactive systems itself remains stable.

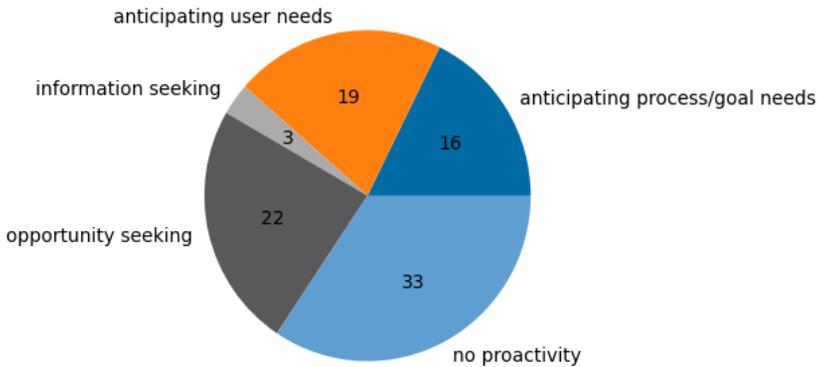

Fig. 8. The distribution of different types of proactive behaviour observed in co-creative systems. The categories include anticipating user needs, anticipating goal needs, information-seeking, and opportunity-seeking behaviours. While some systems exhibit multiple types of proactivity, the overall distribution highlights the prevalence of certain behaviours over others. Duplicate entries from the same paper are included to reflect the total occurrences of each type across the dataset.

The distribution of creative tasks vs proactivity in Fig 14 shows us that game design, music, and performing arts all have a higher percentage of proactive systems. Systems without neural networks ("Other") showed the highest proportion of proactive behaviour.

Figure 15 highlights that game design systems predominantly exhibit behaviours focused on anticipating goal needs. In contrast, systems in the literature tend to prioritize 'anticipate user needs' behaviours, as their primary aim is to assist authors in realizing their creative vision. Music systems, on the other hand, demonstrate a stronger tendency toward opportunity-seeking behaviour, reflecting the improvisational and exploratory nature of music creation. Lastly, visual arts systems primarily focus on anticipating user needs and opportunity-seeking behaviours, with relatively fewer instances of anticipating goal needs.

Despite only five papers focusing on information-seeking behaviour seen in Fig 9, this proactive type covers all phases of the creative process, which is noteworthy. Similarly, although there are few papers focused on the clarification phase of the creative process, it is represented across all types of proactivity, which is also noteworthy. Overall the phases of the creative process against the subcategories of positive proactive behaviour are evenly distributed. Similarly, consistent distribution of Model type categorization can be seen against the subcategories of positive proactive behaviour with neural networks and "Other" type accounting for 40-45% in the subcategories each and LLMs being 14-20%.

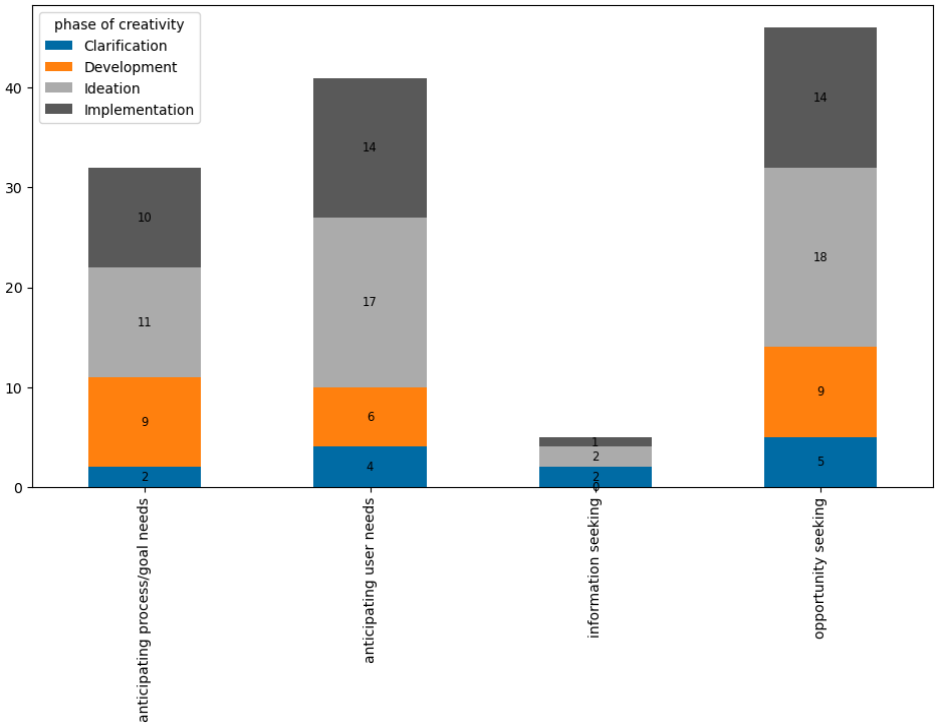

Fig. 9. How different types of proactive behaviour—anticipating user needs, anticipating goal needs, information-seeking, and opportunity-seeking—are distributed across the phases of the creative process.

In the last few categorical dimensions, we looked at what different types of AI techniques could be used to support each subcategory. In the following paragraph, we will attempt to do the same for the types of proactive behaviours AI can show. Here we would like to note that AI techniques alone don't determine the type of proactivity: the system's framing, design decisions, and how it's programmed or prompted play a huge role. Many algorithms are versatile and can support multiple types of proactivity depending on how they're implemented. Such as [57] information-seeking behaviours use cooperative contextual bandits(Other-type model categorization) to see how an image selected by the user fits into the current existing mood board and if it doesn't fit, it asks for justification. However, whether the AI actively seeks information depends on system design: a rule-based framework needs explicit triggers to ask questions, while more adaptive systems rely on dynamic thresholds, like detecting hesitation or inconsistent user input.

Opportunity-seeking behaviour represents the most exploratory type of proactivity, where AI identifies and pursues new directions beyond the user's explicit input. Generative models like GANs or LLMs excel here, as they can produce novel outputs or unexpected ideas. But there are other AItechniques that can also be used like rule-based techniques that can offer alternative recipes possible through switching even a single ingredient or cooking technique in [49] or varying the dance moves a system in [115] uses if the user repeats the same gesture for too long

Anticipating user needs can be supported by all kinds of techniques, with predictive modelling, reinforcement learning or user modelling. [36] uses clustering, some set of rules and a deep Q learning agent to help the user draw by mimicking the user's most recent inputs with some creative adjustments from the system. It even takes into account feedback from the user to learn the needs of the human better. Cooperative contextual bandits, like those used in [57], analyse user inputs such as images on a mood board to identify and suggest additional images that may suit the user's needs for the project. Similarly, systems like [49] combine rule-based approaches and task-specific datasets to offer ingredient suggestions tailored to user preferences during culinary creation. The effectiveness of these techniques depends on system framing, such as when to intervene and how user autonomy is preserved, without the need for complex algorithms. Even rule-based systems can anticipate needs effectively when programmed to monitor specific patterns or contexts.

Anticipating goal needs goes hand in hand with anticipating user needs since the goal for creative systems can often be user-driven or oriented. Occasionally systems offer additional inputs aimed at predicting potential obstacles or requirements and acting in advance to help fix them. Like [10] analyses and uses game definitions/rules to recommend different game mechanics that would suit the game, such as the optimal weapon types to include in the game. These examples show that it's not just the AI techniques but also how they're designed and framed within the system that makes proactive behaviours feel natural, useful, and truly supportive of creativity.

*Noteworthy findings.* User perception of the proactivity of the AI system affecting the experience in a positive way was divided within most papers. Some users appreciated the proactiveness of the system in providing ideas, suggestions or feedback. E.g. - "it helped them understand and reflect on why pictures were chosen" [57]- others still found them frustrating when the AI made changes that went against the vision the users had in mind [39]. This could be circumvented by including the human in the decision-making/application while keeping the system proactive in other aspects as proven by Kraus et al. [60] and Changhoon et al. [85] A suggestion was given by Elgarf et al. [39] in their paper that an adaptive/contextual proactive inputs or interventions may suit users better - "The robot would detect children's frustration at its interference and would only suggest ideas when children are lost or silent for a long time" rather than interrupt a user. However, it has been seen by Kraus et al. [60] that "time-dependent measure for insecurity is insufficient for usage as a trigger-variable of proactive dialogue strategies" So other measures like task difficulty or cognitive-affective states could instead be used as triggers.

The user perception and value of the proactivity within one system could change based on the role in which it was applied as seen with Dang et al.[32], who asked users to rate the value of the proactive ideas of their system. They found that users rated the system's ideas to be useful and that they aligned with their writing. They also reported that users often accepted its suggestions, which supports the finding that the system's proactive idea suggestions were valuable. However, some users also called the suggestions distracting when suggestions were something other than what they had in mind, especially when it came to argumentative writing where they already had opinions. The authors suggested this may have to do with the perception of the role of the system. When the system was seen as a "proposer", a more active collaborator in the process, its proactive suggestions were welcomed. But when it was seen as a "transcriber", a more reactive role, the proactive suggestions were seen as distracting. Winston and Magerko [115] also saw that the humans disliked it when they didn't get to initiate/ lead for the majority of the interaction. This does not mean that the initiative should be human only, but rather that the AI shouldn't be in a leading role. At the end of the paper [115], they suggested shared initiative with a slight bias towards the humans although further research on this may be needed to find out if this design choice would be preferable by users.

Davis et al. [35] reported that the shared initiative in their interactions helped prevent task abandonment. Users, especially novices focused more on maintaining a creative and engaging conversation with the agent over the final results. This can be seen in other papers: The proactive nature of the collaboration in Davis et al. [35] where the users saw the system as a brainstorming partner prompted them to "continue the interaction, which may have helped prevent task-abandonment" and "focused on responding to their partner with an interesting and creative contribution" [35].

Based on this context, we might ask some questions like: Do users have a clear preference regarding the existence of proactivity in a system? Does the framing of the system, such as being perceived as supportive or directive, creative or functional, or an equal collaborator versus a tool, affect their experience? How does this framing shape users' expectations of the system's behaviour? This remains an open area for further exploration, as understanding how these elements interact could clarify when and how proactivity contributes positively or negatively to user experiences.

***Design considerations***. But based on the existing studies done, we get some considerations for designing the proactivity in a system: Effective proactive AI systems should balance support with user autonomy to ***DC 5: provide assistance or structure without undermining the user's control or creative vision*** [60]. The boundaries and limitations imposed by such structures can provide a framework within which creative choices are made. ***DC 6: The system's actions or dialogue should adapt to contextual signals, such as user needs or task states, rather than relying on static triggers*** [60] thus showing a need for "anticipating user needs" type of proactive behaviours. Finally, ***DC 7: the system's role—whether collaborator, assistant, or tool—should be clearly defined*** and ***DC 8: the system's behaviour should be aligned with the system's role*** to gain trust and increase usability.

## 4.4 User Control

The review categorized systems based on levels of user control over the AI, with 28 papers documenting systems where users had no control, 16 papers featuring systems that allowed users limited control, and 18 papers presenting systems where users had full control. In Fig 13 we observe that co-creative systems that are not proactive also tend to have limited or no user control over the AI's contributions.

The degree of user control in co-creative systems varies widely and can be expressed in different ways. In systems with no or low control, users often have no to extremely limited influence over the AI's outputs. For example, in [36], a system where both the human and AI could draw lines, but none could be erased, restricted user control significantly, forcing users to adapt to the AI's contributions. Similarly, improvisational tasks such as those in [24] or [62], often leave users with minimal or no ability to modify the AI's contributions, as the focus lies on real-time interaction rather than end-result customization.

For systems with partial control, users are offered some ability to influence the AI's outputs, though within pre-defined constraints. For instance, systems like [20] present users with system-generated options that can be partially or fully adopted but prevent actions outside those given options. A similar approach is seen in [19], where users rate AI-generated options, guiding the system to create the next iteration based on those ratings but limiting control over the specifics of each new generation. Other diverse modalities of control also exist like in [51], an interactive dance system where users can influence the AI's dance step choices by repeating steps to teach it preferred connections, blending their inputs with the system's autonomy.

At the other end of the spectrum are systems that grant full control, where users retain complete authority over the outcome. In [57], the AI suggests images and tags for a creative mood board, but users can add, replace, modify, or even justify their choices, feeding these decisions back into the system for improved suggestions. Similarly, co-creative songwriting systems [28] allow users to accept, edit, or reject AI suggestions for lyrics and melodies, offering maximum flexibility and customization. Embodied agents, such as the one described in [64], give users the ability to fully physically rearrange blocks added or moved by the AI, thus adjusting the AI's contributions, giving users complete control over the final creative output. This spectrum of user control demonstrates the variety of design choices available, each tailored to different creative contexts and user needs, from guiding improvisational performances to crafting personalized end products.

*Noteworthy findings.* According to Legaspi et al. [65] the more perceived control a human has over a process or end product, the more confident they feel about the outcome. They are also more likely to associate greater authorship with themselves. This may be important to consider when looking at system adoption. Many users in the studies part of this review mentioned wanting more control over the AI or the end product. They were even ready to "sacrifice a certain degree expressivity and exploration of the tool to gain control over the generated content, as the users continuously try to impose their vision" in papers [10] or [58]. This was shown again by [35] where the artists "want to have a means of viewing and manipulating the creative trajectory the system calculates to increase their control over the agent's activities." Koch and his team mentioned that future work "should investigate how to provide more control while keeping the benefits of the AI system." [58]. Others also suggested letting users make most of the decisions to improve their experience, mentioning how this could impact design where "Even if a user receives an order or request from the AI, it might be better to provide him or her with options or ask permission for the request."[85]. [57] proves this with their user study where 8 out of the 16 participants noticed and specifically mentioned that they liked that the AI "left decisions up to them". And again with [70] where having the final editing of the text under their control, the users felt greater agency and ownership over it. This perception of course does not apply to systems where the task IS to leave decisions to the AI system, like improvisational tasks [79], [77] or if the task was designed with the AI making the decisions in mind and conveyed as such beforehand to the participants [43], [34].

**Design considerations**. It's important to give users a real sense of control over the process and outcomes. **DC 9:** *Let users make meaningful decisions to build confidence and help them feel ownership over their work* . **DC 10:** *AI outputs should be transparent and editable* , so users can tweak and refine them to fit their vision. **DC 11:** *Allow users to choose between options suggested by the system or have approval over the system's actions* —this approach keeps the process flexible and user-driven. At the same time, **DC 12:** *balance the user control with the system features that encourage exploration to help users stay creative without feeling overwhelmed* . By putting users in the driver's seat and letting AI act as a helpful partner, these systems can empower people to create with confidence and creative freedom.

## 4.5 System Embodiment

Observations from the system embodiment dimension show that most of the papers in our scope had non-embodied systems(50 papers). The number of embodied systems increased between 2019-2022, although 2023 had none. Most (10 out of the 12) of the embodied systems were improvisation-based task designs, with all the virtually embodied agents belonging to different studies on the same system - LuminAI [101].

Systems designed with embodiment span a diverse range, from physical robots to virtual entities and even non-embodied interfaces. Physical embodiments, like Furhat [39] and Jibo [8], combine the benefits of tangible interaction with features like expressive faces to engage user's creativity. The Yolo [13] system is another type of physical robot used in drawing tasks. Other specialized examples that demonstrate the versatility in physical embodiment include robotic arms, such as those in the Dream Painter project [26], which translate abstract concepts like spoken words into visual art, leveraging embodiment to increase user engagement and interactivity. Cobbie [66] equipped with drawing capabilities and camera input, was designed to explore the integration of co-creative technology into physical robots, enabling collaborative sketching on tangible workstations. Meanwhile, virtually embodied systems like LuminAI in [115] or [62], create engaging collaborative dance experiences through virtual representations. This variety highlights the adaptability of embodiment across contexts, for tailoring interactions in collaborations.

*Noteworthy findings.* Lin et al.[66] specifically tested for the effects of embodiment. They tested embodied vs. non-embodied systems with the same task in a user study. Their research was specifically motivated by the "impact of the embodied robot in providing inspirational stimuli on ideation during conceptual design." They found that their results were consistent with [63] that "robots are more friendly and engaging than non-embodied agents". Lin et al. [66] saw that Cobbie, the embodied agent, was more engaging, and stimulated the user from diverse perspectives due to its multimodal interactions and dynamic movements.

While embodiment can enhance interaction quality and engagement, it also comes with trade-offs. Physical embodiment offers the most tangible and immersive interaction, but it is less flexible and may be impractical in certain settings due to cost and space constraints. Virtually embodied systems allow for creative ways of giving the AI a "body" or a "character," facilitating richer interactions without the logistical challenges of physical robots. However, they may not provide the same level of presence or sensory feedback as physical agents. Non-embodied systems, while highly flexible and accessible, may feel less socially present(discussed further in section 5.1) which could change the engagement levels or quality in certain creative contexts.

**Design considerations**. A more inclusive guideline would be to consider embodiment from both a practical and interaction perspective. ***DC 13: The embodiment of the system should be considered in relation to the task and its interaction demands*** . While embodiment may increase engagement—especially in domains where multimodal stimulation is the norm, such as dance or painting—it may be less critical in other contexts, such as digital art or text-based collaboration. Since multimodal stimulation has been shown to enhance creativity [121], [29], [15], designers should evaluate whether embodiment enhances or hinders the intended interaction.

## 4.6 AI Model Type

To examine the effect of the LLMs entering the field, we look at the distribution of model types by year. Figure 10 shows us that while there is a sizeable jump in the number of papers from 2019 onward, which one may initially correlate to the LLM release timelines [47], they are not solely responsible for the increase in interest in this field. We also see that algorithms without NNs and general NN usages are reducing while LLMs are increasing in co-creativity systems.

Looking into specific types of AI tasks each model was applied to, we observed that Large Language Models (LLMs) were predominantly employed in tasks requiring combined content generation and communication (e.g., [16] and [70]. Neural Networks like Variational Autoencoders (NN-VAEs) are also featured in content generation like in [52]. Generative Adversarial Networks (GANs) were often used for generation tasks or generation combined with communication in prompting to generate outputs/ideas, while traditional convolutional/recurrent neural networks (CNNs/RNNs) supported a mix of classification, content generation, and decision-making roles, particularly when integrated with heuristic or geometric methods (e.g-[35]). In contrast, symbolic reasoning and code-based AI models excelled in personalization, optimization, and knowledge retrieval tasks. For instance, "Recipe 2.0" [49] applied such models for culinary idea generation through personalized recommendations, while "May AI?" [57] used non-NN models in optimization and knowledge curation. These models can contribute to communication parts of the task, as seen in "Once Upon a Story"[39]. And, some machine learning models can even do content generation [116], [18]. This categorization is relevant because it shows how different AI models are tailored to specific types of tasks, offering insights into their strengths and potential applications in diverse creative and problem-solving contexts.

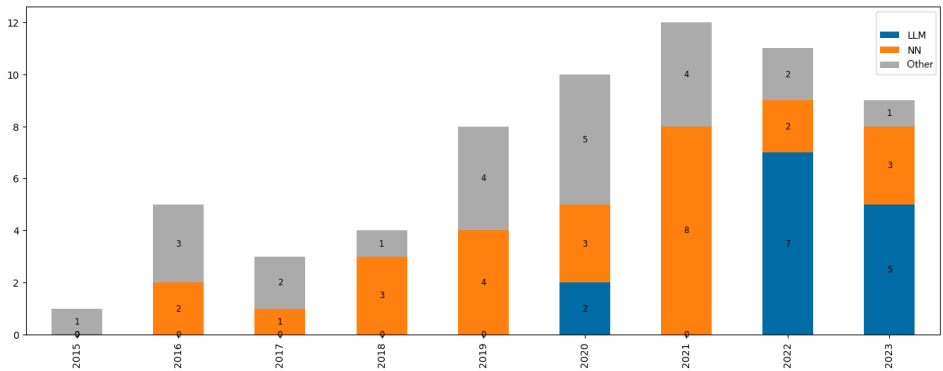

Fig. 10. The distribution of different model types—large language models (LLMs), general neural networks (NNs), and non-neural approaches—across publication years. While there is a notable increase in the number of papers from 2019 onward, this growth is not solely attributable to the emergence of LLMs.

## 5 Evaluation Themes

While the previous sections focus on dimensions central to the design of co-creative systems, the evaluation themes section provides additional insights into how users interact with these systems. This section explores these broader themes, including perceptions of system intent, role expectations, transparency, and consistency. These findings provide a richer understanding of the challenges users face, the strategies that enhance engagement and trust, and the implications for designing systems that truly empower creativity.

### 5.1 User Perceptions of System Intent

Several results in the surveyed papers highlighted the anthropomorphisation tendencies in users. [57] and [108] both had half of their participants "perceive and consider the agent a teammate" even though it was known that the agent was artificial. They noted that "its identity, especially its name fostered the perception of a social artificial teammate"[108]. This association of agency to the AI system has been shown to affect humans' trust in the system [76] and the overall experience. Since this tendency of anthropomorphisation is a fundamental and widely prevalent cognitive behaviour [7] that is also seen in human-AI interactions [76], accounting for this phenomenon should be a baseline consideration in system design. We lay out some examples of this phenomenon occurring in the papers in the scope of this review, and what effects it had on the user experience to help anticipate how system interactions may play out.

Humans attribute mental and affective states when they anthropomorphize non-humans [7]. This caused a noticeable association of social meaning to the interaction with the AI, irrespective of whether it was designed to elicit such meaning. This "projected inference" [76] of meaning in the actions of the AI system was seen in many papers like [66] or in [108] where the system was built with a "waiting period" before giving the contribution to the project causing the users to "perceive it as a socially present teammate thinking about its next contribution." When comparing systems with and without this waiting period they noted that without it the system was NOT perceived to be thinking and therefore less socially present.[108].

Such meaning could have a neutral or positive emotional effect or value in the interaction. [115] had participants associate colours in the system with "thinking mode" or "hesitation" for red and "understood signal" or "confidence" for blue, even though the intended meaning of these colours was about a different functionality (lead vs follow role of the AI). In the case of [35] where participants, especially novice users of the system, would stop worrying about the final outcome of the task and become more focused on responding to their "partner" with an "interesting and creative contribution" which helped prevent task abandonment.

However, there is always the risk that the social meaning may end up being negative. [64] had a robot with no social behaviours or meanings planned as part of its interaction design. This robot during its interactions with the users would sometimes reverse the users' ideas without giving any reasoning from its side. This caused a social interpretation of "dismissiveness" to the user. Similarly in [115] found that in the lead mode, the system was disliked by some users due to the feeling of being "ignored". [8] was found to be "rude" by participants as it would often not respond or interrupt. The instances it took a while to respond were perceived as "not listening" to the users. In [108], one participant mentioned that they felt sorry for the agent when taking over and editing/changing a sentence the system had written, although most of the other participants found it easier to edit the system's sentence over a follow human teammate's.

**Design considerations**. Based on these accounts and the other studies done on the phenomenon, we lay out some design considerations to thoughtfully incorporate elements that leverage this behaviour to enhance user experience while minimizing potential negative effects. For example, one can **DC 14:** *encourage users to perceive the system as a socially present collaborator rather than a tool* by using intentional pauses or delays in system responses to simulate thoughtfulness. This social perception of the AI system has been shown to increase trust in the system by the users, which is essential in co-creative settings where collaboration hinges on mutual understanding and shared contributions [76]. Similarly, use visual or auditory cues, such as colours or tones to signal internal system states (e.g., "thinking" vs. "ready"), to inculcate a sense of collaboration and understanding, provided these signals are intuitive and clearly aligned with their intended purpose [117]. **DC 15:** *Pilot user studies should be used to evaluate the impact of anthropomorphic features, such as tone, timing, and language used by the system, on user perceptions and outcomes*, for finer refinements of study context-specific interactions. **DC 16:** *Assign the system a name or identity to enhance its perceived social presence, but care should be taken to ensure that this identity aligns with its role to avoid setting unrealistic expectations*. It is also important to mitigate behaviours that could lead to negative associations. For example, ambiguous or unexplained system actions, specifically if they alter a contribution from the user, may be perceived as dismissive or rude and disrupt the trust and collaboration dynamics as well as stifle creativity by limiting the flow of ideas. Communicating the rationale behind these behaviours and providing timely feedback can help address this. In these ways, by anticipating and accommodating anthropomorphic attributions designers can create systems that are more intuitive, trustworthy, and effective in supporting users. In the context of co-creativity, these considerations not only improve the user experience but also create an environment that nurtures creative exploration.

### 5.2 Roles and Expectations

Ryssina et al. [96] talk about how the distribution of roles of members within a team based on various factors like expertise, team structure and interpersonal relations can affect the creative outcome of the task at hand. "They found that which roles were the key depended on the stage of progress of a project." [96] Therefore negotiation of roles at either the beginning of a collaboration or throughout it may be beneficial for the creative process. Guzdial and his team [43] found in their analysis the four major roles the humans expected the AI to take on, in their system, - friend, collaborator, student, and manager. They also noticed that -"These expected roles could positively or negatively impact the user experience, and fluidly changed throughout each design session." [64] and [10] also remarked on the fact that the users actively took advantage of the flexibility of the roles between the humans and the computer, enhancing and simplifying the experience. [58] had 30% of participants mention them noticing "that each designer's role in ideating or searching for images switched throughout the process: one guides and the other follows, but then that changes through time depending on what you like in different states." While dynamic negotiation of the roles was not happening in every system, the systems were still put into different roles by different users.

An important finding of Ryssina's. [96] study into role structures and creative potential of working teams, was that clarity in the perception of the roles by the driver of the creative process was a powerful influence on the effectiveness of the team/ collaboration. We look into some examples of how the roles were perceived by the participants of the user studies in the papers we reviewed and what factors affected this perception of roles. [32] found that "Some people clearly saw the system as something that serves input efficiency (i.e. transcriber ), whereas others saw it as providing inspiration (i.e. proposer)." Often age, job/background or even experience with the system can affect the roles users associate with the system. In [119] younger children tended to see the system as a partner while older children treated it like a tool. [35] found that designers would think of how the system "could perform the role of their partner so they may engage in collaborative brainstorming" while artists focused more on how it "might help them draw better and 'get things done.'".

It is important to note that once again the smallest system or interaction framing choices can also influence the role perception of the AI or AI features by the users as seen in the following examples. [31] had a summarization system, which was meant to prompt reflection on the user's writing, and had a "Replace in Text" button next to the summaries. "This partly made people perceive summaries as text suggestions that should be included in the text, rather than annotations", thus casting the AI in a different role than initially intended. Upon replacing the button with a "Copy to Clipboard" one, the AI was placed in a more "passive" role, which was corroborated by their second study. Karimi et al. [54] had a study which also found that the "low novelty" setting in their system was used by the designers in the finishing touches/ adding more details. This was a different role from that of the "high novelty" mode which was used to transform or add to the existing design.

The nature of the work makes a difference in the user preferences of roles. In improvisational tasks, like dancing with the system, Several participants in [115] and [62] preferred the agent in the "follow" mode due to more perceived responsiveness to the user's contributions. They also cited mimicry from the AI as being a preferred response. [59] shows that humans prefer taking the leader role in the majority of the interactions, and the same was concluded by [115]. This may be limited to improvisation since [59] was also a drumming game improvisational task. [85], a drawing task-based study noted that users liked to make most of the decisions and take the initiative in collaborating with the system in the experiments although in the surveys this preference was not very significant. "Users' ability to make the decision at every moment seemed more important than being in the Lead mode itself" [85] Thus maybe it is less about the role of leading and more about the level of control over the creative design/ result that is more important to the users in non-improvisational systems. Further research may be required on this.

The role of AI in collaborative systems is closely tied to dimensions studied in previous sections like the proactivity of the AI, user control, and the phase of the creative process targeted by the system. AI's proactivity can determine its role, with anticipatory behaviours casting it as a proactive assistant or collaborator, and opportunity-seeking actions often positioning it as a creative partner. In contrast, reactive behaviours align AI with tool-like roles. User control is another critical factor, as high user control positions the AI as a tool or assistant, while shared control garners a partner dynamic, and low control often casts the AI as a leader or manager. The phase of the creative process can also influence role negotiation: during ideation, users may welcome an AI collaborator providing diverse suggestions, while in the refinement stages, they often prefer the AI as an assistant focused on detail-oriented tasks.

**Design considerations**. To effectively support co-creativity, systems should dynamically adapt to users' evolving needs and preferences, by **DC 17: *enabling fluid role shifts that align with user preferences and task needs*** for productive and satisfying collaborations. For example, incorporating explicit role negotiation features, like toggles for "high novelty" vs. "low novelty" modes as seen in Karimi et al. [54], can help users adapt the AI's role based on their needs for exploration or refinement in the ongoing phase of the creative process. Additionally, **DC 18: *designers should consider user demographics and experience levels when designing personalized system collaboration dynamics***. For instance, novices or younger users are more likely to put the system in a partner-like role, for they may benefit from systems encouraging creative experimentation and learning through collaborative brainstorming. In contrast, experienced professionals were shown to view the system as a tool, so they might benefit from being provided with targeted suggestions or solutions without interrupting their creative flow.

### 5.3 System Transparency and User Guidance

Around 30% of the papers in our review had comments about instructions on how the system worked and what was expected of the users. Participants often mentioned being unclear or confused about how their actions affected the system, especially their feedback throughout the process in [36], [100] and [34]. In some studies, this miscommunication caused participants to try and teach the agent through other methods, like repetition, which the system wasn't designed for. This mismatch of expectations can cause frustrations in the overall experience [34]. "Miscommunication of the design goals for the installation - it became clear that some users felt like they should have been able to control the agent's actions to a greater degree." [51]. [115] and [62] also had this issue where the change from "follow" to "lead" mode communicated through changing the lights from blue to red, was not noticed by users or misunderstood as the mental states of the agent. This also caused the agent to be disliked due to seemingly "ignoring" the user "randomly" (it had switched to the lead role from the following role) since the change in role was not clear to the users. "The discrepancy between user perception of the AI and how AI actually works significantly compromises the usability of the system" [106]. Finally, such miscommunications can also cause confusion regarding the expected role of the AI [31] or the role users themselves. As with [26] where the "spectators wait for the technology to be very creative and magical while undermining their own role."

When detailed instructions were given, participants noted and appreciated it. "It improved users' perceived predictability, comprehensibility, and controllability of the system and task" [85], [11] and [70]. Such instructions need not be a large wall of text given at the beginning of the study or interaction. [32] offered support by way of examples for the feature. [108] had the agent self-introduce. They saw that this self-introduction caused "most of the participants to have realistic expectations toward it" which even prompted a more forgiving outlook towards the mistakes of the agent. In fact, it was found in a few other studies too, that if the users could understand why the AI did what it did, its contributions were more acceptable, even if the AI had altered what the user had done [11], [36], which had initially caused frustration "if it 'messed up' something they drew". [57] also noted that the "passive explanation" and the "proactive questioning feature" helped users understand and reflect on why the pictures were chosen/ suggested.

The most important part of this need for transparency is that it builds trust towards the AI as a collaborator. They saw conflicts arise when "users lost confidence in the robot's strategy or felt that the robot was not being mindful of theirs." [64]. Users said they need to understand the AI's strategies to work with it. They even felt frustrated when they "could not determine the robot's strategy". "While it may not be necessary for the robot and human to share the same implicit preferences about common design goals, it is important that they are aware of each others' preferences and how they evolve, in order to negotiate them together." [64]. And finally, in [61] saw that the users enjoyed "the feeling of being seen or understood by the system", as it can increase trust in the AI. This is especially important in co-creative systems where seeing the system as a partner/collaborator is important.

***Design considerations.*** Effective co-creative systems thrive on clearly defined and transparent roles that align user and AI expectations, ensuring smooth collaboration. Ambiguous or subtle role cues should be avoided; instead, ***DC 19:*** *the system should provide explicit descriptions or cues for each participant's role or shift between roles to prevent misunderstandings* that could disrupt the creative flow. For example, dynamically adapting instructions—such as through agent self-introductions, contextual prompts, or interactive help—can ***DC 20:*** *make it easier for users to grasp the system's functionality without overwhelming them with dense upfront explanations*.

***DC 21:*** *The system should also clarify its behaviour, particularly in instances of unexpected actions*. This clarity reassures users, helps them understand how the AI's contributions fit into their creative process and reduces frustration. Finally, a sense of partnership is key in co-creativity, therefore the system can simulate this by ***DC 22:*** *demonstrating an awareness of user actions and goals* and enabling collaboration through shared or negotiated preferences.

### 5.4 System Consistency

One common theme that emerged out of the user evaluations of the papers in our scope was consistency, whether it was about AI behaviour or the AI's contributions to the end result. When inconsistency occurs in the creative outputs from the AI, it can cause a break in user engagement and other frustrations in the overall experience. For example, [11] found that the abstractness of the quest actions suggested by the AI caused a "lack of thematic and concrete elements such as NPC roles or defined plot lines and plot elements to follow." This made it harder for the designers to "contextualize their creations", reducing the tool's perceived usability. [17] also had similar issues with consistency in the world-building and quest descriptions, reducing player engagement and overall satisfaction. [61] had participants who commented on the "occasional lack of consistency as a detriment to the overall storytelling experience". This issue also applies to other domains like visual arts, where users said they "found it strange when there was a mixture of high and low-quality objects on the canvas." [85]

AI behaviour inconsistencies cause an equally unusual problem. Participants in [85] felt "uncomfortable" when the system shifted between human-like and non-human-like behaviours. The cognitive dissonance this may cause in the users can feel exhausting and frustrating. In the same experiment, the users also mentioned feeling "confused when the ability of the AI differed across functions". "They felt frustrated when the AI showed human-like characteristics and machine-like characteristics in the same task"

***Design considerations.*** By ensuring consistency and clearly communicating the system's capabilities, co-creative tools can better support the iterative and collaborative nature of creative tasks, allowing users to focus on exploration and refinement without being hindered by unpredictability or frustration. To address this, the system design should **DC 23: *prioritize uniformity in system outputs, ensuring coherence and quality across contributions*** as it increases trust in the system [95]. For instance, tools generating creative content, such as quest actions or world-building elements, should include mechanisms to maintain narrative continuity and logical connections between elements. This could involve structured prompts or templates to ensure all outputs, like NPC roles or plot elements, are fully realized and contextually relevant, reducing user frustration with abstract or incomplete suggestions. Similarly, in visual arts applications, ensuring consistent quality across generated objects—such as avoiding mismatched styles or blending low-quality and high-quality assets. If different features in a single system exist that may have different capability or quality levels, then providing users with upfront expectations of the AI's role and function, coupled with transparent feedback mechanisms for any discrepancies, can also mitigate frustrations when the AI demonstrates uneven proficiency across features. **DC 24: *Different names(identities) and roles may also be given to different features to disassociate them from each other as intelligent entities, causing them to be perceived as different agents with varying levels of capability as compared to one inconsistent and untrustworthy or frustrating agent*** .

## 6 Conclusion

This systematic review of 62 papers on human-AI co-creativity has identified key dimensions, trends, and design considerations that shape the development of collaborative systems in creative domains, providing valuable insights into how these systems can support and enhance the creative process. Our findings highlight the importance of system embodiment, proactive behaviours, model types, and user control in creating meaningful human-AI collaborations. Below, we reflect on the broader implications of these findings and suggest directions for future research.

Certain stages of creativity, such as problem clarification, remain under-explored. Similarly, while ideation-focused systems dominate the field, there is growing interest in supporting development and implementation phases. Embodied systems showed promise for multimodal tasks and enhanced user engagement, while Large Language Models (LLMs) expanded the scope of applications in literature and content generation. However, user studies revealed that effective collaboration hinges on careful role assignment, with high user control often leading to greater satisfaction, trust, and a sense of ownership over creative outputs. Users are more likely to embrace systems that anticipate their needs and offer support without undermining control or creative direction. Systems that exhibited proactive behaviours demonstrated potential for deeper collaboration, though excessive proactivity or poorly timed interventions frustrated users, emphasizing the need for adaptive and context-aware designs. The role of proactivity in a system remains a subject of ongoing exploration. Whether users prefer proactivity depends largely on how the system's behaviour is framed—whether it is seen as supportive or directive, creative or functional, or as a partner versus a tool. Understanding these dynamics will be critical in designing systems that balance AI initiative with user autonomy.

Our findings also point to several important design considerations. Encouraging users to externalize their thoughts—whether through detailed inputs or by comparing their ideas with AI outputs—makes reflection a central part of the creative process. Systems that introduce task shifts, such as moving from storytelling to prompt crafting, help users view their work from new perspectives, fostering deeper engagement and creativity. Creating moments of productive tension, such as reconciling mental summaries with AI-generated ones, can prompt users to engage more deeply with their creative ideas and refine them.

A sense of partnership between users and AI can be enhanced by simulating social presence, for instance, by assigning the AI a name or identity that aligns with its role, increasing trust without setting unrealistic expectations. Additionally, providing dynamic, context-sensitive instructions—such as self-introductions by the agent or examples of system capabilities—can help users understand the system's functions without overwhelming them with dense, upfront explanations. Systems with multiple features of varying capabilities should clearly communicate their roles and functions upfront, providing transparent feedback mechanisms to manage user expectations and avoid frustration when AI demonstrates uneven proficiency across different features.

As co-creativity systems advance, ethical challenges surrounding intellectual property and the consensual use of artist data should be addressed in any paper studying them. Many AI models are trained on vast datasets, which may include copyrighted works or other protected materials. This raises concerns about the ownership of AI-generated content, particularly in creative fields where the boundaries between human and machine contributions blur. Future systems should prioritize transparency, ensuring that artist data is used consensually and that AI-generated outputs provide clear attributions where applicable.

Despite the progress made, several limitations remain in current research. Systems focusing on the early phases of creativity, such as problem clarification, are under-represented, and evaluation methods often rely on subjective metrics like perceived creativity, limiting our understanding of long-term impacts on collaborative dynamics. Future research should explore adaptive system designs that allow users to shift roles fluidly during collaboration, as well as examine how roles evolve over time, based on user preferences and task requirements. Proactivity should also be designed to be context-sensitive, with AI systems anticipating user needs or providing task-specific interventions to enhance collaboration without undermining user autonomy. Ensuring that AI outputs are transparent and editable fosters a stronger sense of authorship and trust, particularly in non-improvisational tasks where users desire control over the final creative product.

In addition to these challenges, the adaptability of users and the learning curve associated with AI tools pose further limitations. Many users, particularly those without prior experience with digital or AI-driven systems, may struggle with initial interactions, hindering accessibility and limiting the system's broader adoption. Ensuring that these systems are designed with intuitive interfaces and sufficient guidance will be critical in enabling a wider range of users—both novices and experts—to effectively collaborate with AI in creative tasks. Improving user accessibility will be instrumental for the continued growth and inclusivity of human-AI co-creativity.

Human-AI co-creativity holds the potential to enrich creative practices by offering new tools and approaches for collaboration. By addressing current limitations and ethical concerns, future co-creative technologies can foster more meaningful and productive collaborations that respect the rights and contributions of all creative agents involved.

## Acknowledgments

This research was funded by the Hybrid Intelligence Center, a 10-year programme funded by the Dutch Ministry of Education, Culture and Science through the Netherlands Organisation for Scientific Research, https://hybridintelligencecentre.nl, grant number 024.004.022.

A special thanks to Imaginaolgy Quantum Craft Laboratory for granting me access to invaluable resources and insights on human-AI co-creative artists. Their knowledge and collaboration have greatly enriched my understanding of this field. Finally, I acknowledge the use of ChatGPT (OpenAI, https://chatgpt.com/) to help rephrase and format my text and to proofread my final draft.

A Systematic Review of Human-AI Co-Creativity                                                                                                   x:31

## Appendix

The following appendix presents a set of images and a table that visually and systematically supports the findings discussed in this review. These visuals illustrate key trends, system categorizations, and design considerations identified in the surveyed literature on human-AI co-creativity. Each figure provides a structured representation of the data, offering insights into the distribution of creative tasks, system embodiment, proactive behaviour, and user control, among other dimensions. The accompanying table consolidates the categorization of co-creative systems, summarizing the relationships between different design choices and their impact on user interaction.

The figures and table are derived from the systematic analysis of 62 papers and are organized to align with the dimensions outlined in Section 4 and the evaluation themes discussed in Section 5. Readers are encouraged to refer to the figure captions and table descriptions for a detailed explanation of each visualization and its relevance to the analysis. Together, these materials serve to complement the textual findings, reinforcing the broader discussion on how different design choices shape co-creative interactions.

## A

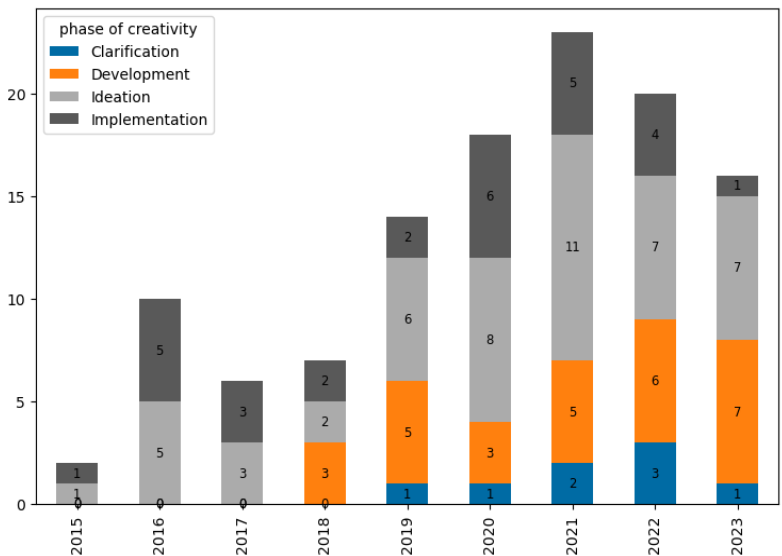

Fig. 11. The distribution of studies focusing on different phases of the creative process—Clarification, Development, Ideation, and Implementation—across publication years from 2015 to 2023.

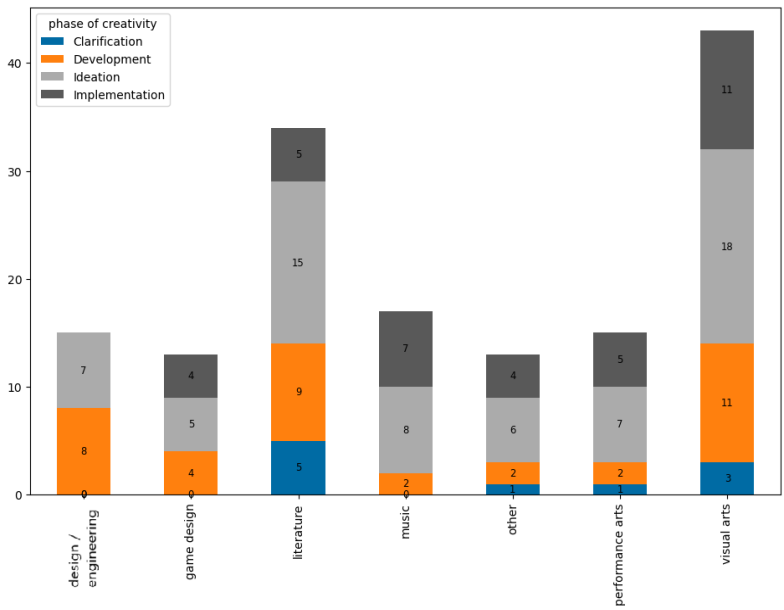

Fig. 12. How different creative domains—such as Visual Arts, Literature, Music, Performing Arts, Design, and Engineering—are distributed across the phases of the creative process.

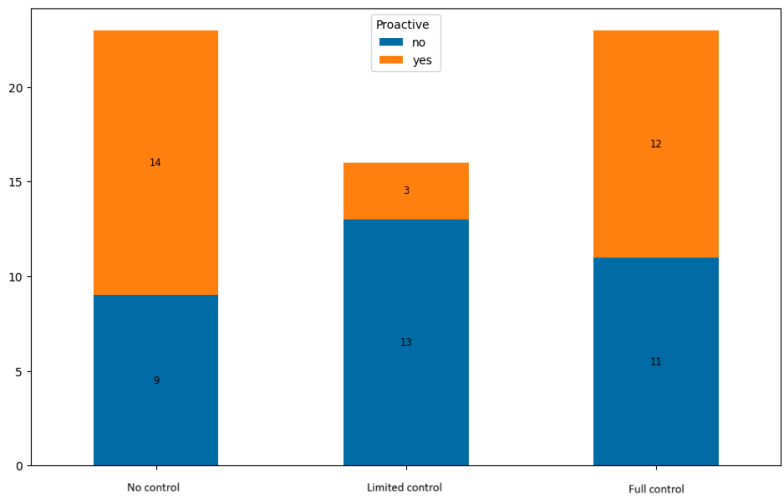

Fig. 13. The relationship between system proactiveness and the degree of user control over AI contributions.

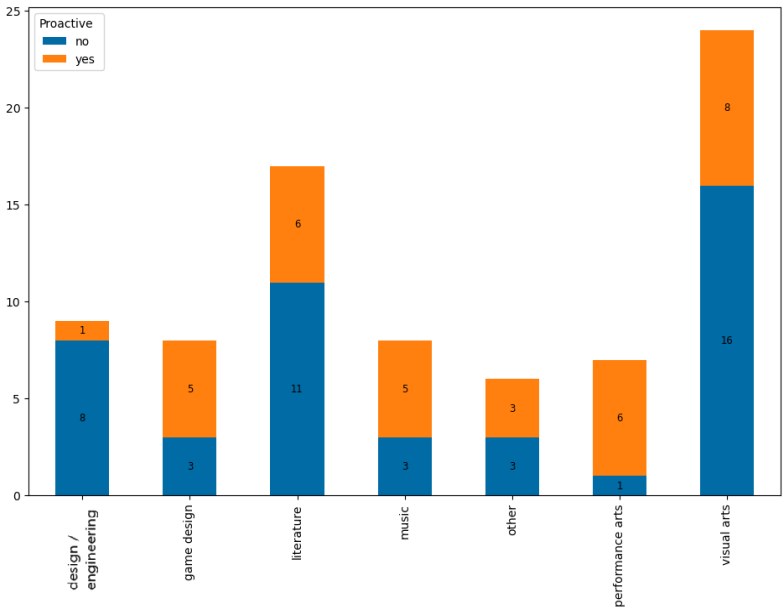

Fig. 14. The distribution of system proactiveness across different creative domains.

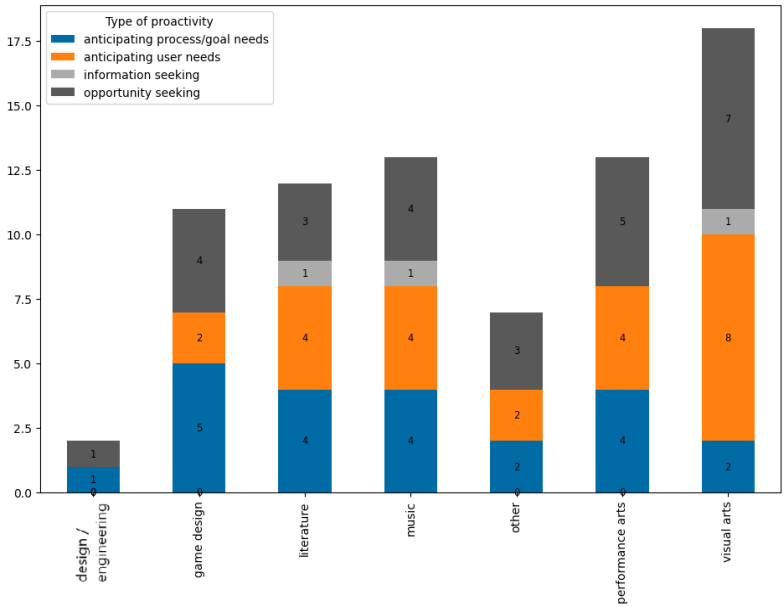

Fig. 15. How different creative domains emphasize various proactive behaviours.

## B

| Design consideration | Description |
|---|---|
| **Creative Task** | |
| DC 1 | Encourage users to explore unconventional connections between creative domains and develop hybrid forms of expression, pushing beyond the patterns that exist within any single domain. |
| **The Phase of the Creative Process** | |
| DC 2 | Introduce task shifts to help users see their work from fresh angles and spark deeper thinking. |
| DC 3 | Encourage users to externalize their thoughts to make reflection a central part of the creative process. |
| DC 4 | Create moments of productive tension to push users to engage more deeply and refine their creative ideas. |
| **Proactive Behaviours of the System** | |
| DC 5 | Provide assistance or structure without undermining the user's control or creative vision. |
| DC 6 | The system's actions or dialogue should adapt to contextual signals, such as user needs or task states, rather than relying on static triggers. |
| DC 7 | The system's role—whether collaborator, assistant, or tool—should be clearly defined. |
| DC 8 | The system's behaviour should be aligned with the system's role. |
| **User Control** | |
| DC 9 | Let users make meaningful decisions to build confidence and help them feel ownership over their work. |
| DC 10 | AI outputs should be transparent and editable. |
| DC 11 | Allow users to choose between options suggested by the system or have approval over the system's actions. |
| DC 12 | Balance the user control with the system features that encourage exploration to help users stay creative without feeling overwhelmed. |
| **System Embodiment** | |
| DC 13 | The embodiment of the system should be considered in relation to the task and its interaction demands. |
| **User Perception** | |
| DC 14 | Encourage users to perceive the system as a socially present collaborator rather than a tool. |
| DC 15 | Pilot user studies should be used to evaluate the impact of anthropomorphic features, such as tone, timing, and language used by the system, on user perceptions and outcomes. |
| DC 16 | Assign the system a name or identity to enhance its perceived social presence, but care should be taken to ensure that this identity aligns with its role to avoid setting unrealistic expectations. |
| **Roles and Expectations** | |
| DC 17 | Enable fluid role shifts that align with user preferences and task needs. |
| DC 18 | Designers should consider user demographics and experience levels when designing personalized system collaboration dynamics. |
| **Transparency** | |
| DC 19 | The system should provide explicit descriptions or cues for each participant's role or shift between roles to prevent misunderstandings. |
| DC 20 | Make it easier for users to grasp the system's functionality without overwhelming them with dense upfront explanations. |
| DC 21 | The system should also clarify its behaviour, particularly in instances of unexpected actions. |
| DC 22 | Demonstrate an awareness of user actions and goals for a sense of partnership. |
| **Consistency** | |
| DC 23 | Prioritize uniformity in system outputs, ensuring coherence and quality across contributions. |
| DC 24 | Different names (identities) and roles may also be given to different features to disassociate them from each other as intelligent entities, causing them to be perceived as different agents with varying levels of capability as compared to one inconsistent and untrustworthy primary agent. |

Table 3. List of the final 24 design considerations synthesized from results and learnings of the 62 papers reviewed